\documentclass[pra,twocolumn,aps,superscriptaddress]{revtex4}
\usepackage{latexsym,graphicx}
\usepackage{amsmath,amssymb}
\usepackage{epsfig}
\begin{document}

\title{\large\bf{Landau damping: Instability mechanism of superfluid Bose gases moving in optical lattices}}
\author{Kiyohito Iigaya}
\email[email]{:iigaya@kh.phys.waseda.ac.jp}
\affiliation{Department of Physics, School of Science and Engineering, Waseda University, Okubo, Shinjuku-ku, Tokyo 169-8555, Japan}
\author{Satoru Konabe}
\affiliation{Department of Physics, Faculty of Science, Tokyo University of Science, Kagurazaka, Shinjuku-ku, Tokyo 162-8601, Japan}
\author{Ippei Danshita}
\affiliation{Department of Physics, School of Science and Engineering, Waseda University, Okubo, Shinjuku-ku, Tokyo 169-8555, Japan}
\affiliation{Physics Laboratory, National Institute of Standards and Technology, Technology Administration, U.S. Department of Commerce, Gaithersburg, Maryland 20899-8410, USA}
\author{Tetsuro Nikuni}
\affiliation{Department of Physics, Faculty of Science, Tokyo University of Science, Kagurazaka, Shinjuku-ku, Tokyo 162-8601, Japan}
\date{\today}
\begin{abstract}
We investigate Landau damping of Bogoliubov excitations in a dilute Bose gas moving in an optical lattice at finite temperatures. Using a 1D tight-binding model, we explicitly obtain the Landau damping rate, the sign of which determines the stability of the condensate. We find that the sign changes at a certain condensate velocity, which is exactly the same as the critical velocity determined by the Landau criterion of superfluidity. This coincidence of the critical velocities reveals the microscopic mechanism of the Landau instability. This instability mechanism shows that a thermal cloud plays a crucial role in breakdown of superfluids, since the thermal cloud is a vital source of Landau damping. We also examine the possibility of simultaneous disappearance of all damping processes.
\end{abstract}

\maketitle
\section{\label{intro}introduction}

Superfluidity, one of the most startling phenomena in atomic and condensed matter physics, has fascinated both experimentalists and theorists since its first discovery in liquid Helium \cite{Kapitza, jack allen}. Unfortunately, the inter-atomic interaction of liquid Helium was too strong to investigate the microscopic properties of superfluidity. However, the revolutionary realization of the Bose-Einstein condensation (BEC) in alkali gases \cite{Anderson, Bradly, Davis} provided us new fertile grounds for microscopic investigations. The diluteness of the gases allows us to treat many problems, including the static and dynamical properties of the BEC, in a microscopic way \cite{PandS}.  
In fact, a number of captivating properties of superfluidity have been revealed in dilute Bose gases such as quantized vortices \cite{Madison, Hodby, Feder} and the Josephson effect \cite{CataliottiJ, Albiez}. Moreover, with the recent technological advancement in optical laser physics, several remarkable experiments on the breakdown of superfluidity in optical lattices have been reported \cite{Burger,Cataliotti,Fallani, Sarlo, Morsch}. 

Regarding the breakdown of superfluidity, it has been known that a superfluid current has its unique critical velocity, above which the current suffers friction, leading to the decay of superfluidity \cite{Kapitza, Dash, Raman}. Landau pointed out that this decay is due to the appearance of negative excitation energy, inducing the energetic instability of the superfluid state \cite{Landau}. Applying Landau's argument to the uniform system of a dilute Bose gas, the critical velocity is known to coincide with the Bogoliubov sound velocity \cite{PandS}. Although Landau originary assumed the homogeneity of the system, his idea of the energetic instability, which is known as {\it Landau instability}, is also applicable to non-uniform systems such as the BEC in optical lattices \cite{Niu, Menotti}.

In addition to the Landau instability, another kind of instability is passionately studied in recent years \cite{Fallani, Sarlo, Niu, Menotti, Demler}. It is called {\it dynamical instability}, which is theoretically related to the appearance of the imaginary component of excitation energy \cite{Niu, Menotti}, indicating the exponential increase of collective oscillations of the condensate. Thus the dynamical instability is expected to be observed even at $T=0$, while the Landau instability is not likely to be observed at $T=0$, since it requires the thermal (non-condensate) component that receives the energy emitted by the condensate during the breakdown process. 

These two kinds of instabilities were experimentally studied for Bose gases in moving optical lattices in Refs.~\cite{Fallani, Sarlo}. Their results show that the energetic instability takes place only at finite temperatures with a sufficient amount of thermally excited non-condensate atoms. This experimental result strongly suggests that thermal clouds are crucial to induce the energetic instability of the superfluid flow. 

In this paper we theoretically investigate the effect of the thermal clouds on the stability of the superfluid current. It should be noted here that this type of issue cannot be studied by the usual Gross-Pitaevskii equation containing no thermal component. 

We restrict ourselves to the collisionless regime. In a low density system at a low temperature, the mean free path of the excitations is known to become comparable with the size of the system itself and collisions between atoms play a minor role \cite{Giorgini}. 

Recently, Konabe and Nikuni \cite{Konabe} studied the instability of the condensate with a superfluid current in a 1D optical lattice using a dissipative Gross-Pitaevskii equation \cite{ZNG} in the collisionless regime. They showed that the collisional damping of a collective mode due to the collisions between the condensate and the thermal component leads to the instability at a critical quasimomentum of the condensate, which exactly coincides with the Landau criterion.

In contrast, in this paper we focus on the role of Landau damping process in the instability of the condensate. Landau damping arises due to a coupling between a collective mode of the condensate and a thermal excitation. This damping process is known to be a dominant damping process in the collisionless regime in the usual trapped Bose gas. In fact, Williams and Griffin \cite{Williams} showed that the Landau damping rate is larger than the collisional damping rate, which is studied in Ref.~\cite{Konabe}. Landau damping has been studied by many theorists in uniform \cite{HM, SK}, trapped \cite{Liu, Pita1, Fed, Giorgini, Guilleumas, Jackson}, and optical lattice \cite{Tsuchiya} systems. While all of these researches focused on current-free systems, in this present paper we study the Landau damping in a current-carrying dilute Bose gas and discuss the instability of the Bose-Einstein condensate induced by the Landau damping process. 


This paper is organized as follows. In Sec.~\ref{LD}, we develop the finite-temperature mean-field theory and derive the expression of the Landau damping rate for a current-carrying Bose gas. In Sec.~\ref{ft-tb}, we develop the mean-field theory in a moving optical lattice within the tight-binding approximation. In Sec.~\ref{ld1}, we obtain an expression for the damping rate for a Bose gas in a moving optical lattice, and we also examine the disappearance of all the Landau damping processes. In Sec.~\ref{LL}, we analyze the derived damping rate and discuss the stability of the condensate induced by the Landau damping process, where we find that the Landau damping process is vital for the Landau instability. In Sec.~\ref{inst}, we discuss the similarity and difference between the Landau instability and the dynamical instability.

\section{\label{LD}finite-temperature mean-field theory and landau damping}

In this section, we develop a finite temperature mean-field theory \cite{Griffin} and introduce the Landau damping rate for a current-carrying Bose gas. The ground canonical Hamiltonian is given by
\begin{eqnarray}
\hat{K}&=&\hat{H}-\mu\hat{N} \nonumber \\  
       &=&\int d{\bf r}  \hat{\psi}^{\dagger}({\bf r},t)\Bigl(-\frac{\hbar^2\nabla^2}{2m}+V_{\rm ext}-\mu\Bigr)\hat{\psi}({\bf r},t) \nonumber \\
 & & {}+\frac{g}{2}\int d{\bf r} \hat{\psi}^{\dagger}({\bf r},t)\hat{\psi}^{\dagger}({\bf r},t)\hat{\psi}({\bf r},t)\hat{\psi}({\bf r},t) ,
\end{eqnarray}
where $V_{\rm ext}$ is an external potential, $g=\frac{4\pi\hbar^2a}{m}$, $a$ is the s-wave scattering length, and $m$ is the atomic mass. The Heisenberg equation for the Bose field operator $\hat{\psi}({\bf r},t)$, which satisfies the Bose commutation relation, is then 

\begin{eqnarray}
i\hbar\frac{\partial \hat{\psi}({\bf r},t)}{\partial t}&=&\Bigl(-\frac{\hbar^2\nabla^2}{2m}+V_{\rm ext}-\mu\Bigr)\hat{\psi}({\bf r},t) \nonumber \\
           & & {}+g \hat{\psi}^{\dagger}({\bf r},t)\hat{\psi}({\bf r},t)\hat{\psi}({\bf r},t) .
\label{EQM1}
\end{eqnarray}
Separating out the condensate part from the non-condensate part in the usual manner, the field operator is expressed as

\begin{eqnarray}
\hat{\psi}({\bf r},t)=\Phi_0({\bf r})+\tilde{\psi}({\bf r},t),
\label{ansatz1}
\end{eqnarray}
where $\Phi_0({\bf r})=\langle\hat{\psi}({\bf r},t)\rangle$ is the time-independent, spatially inhomogeneous order parameter. Substituting the ansatz (\ref{ansatz1}) to Eq.~(\ref{EQM1}), and using the mean field approximation, we obtain

\begin{eqnarray}
\left[ -\frac{\hbar^2\nabla^2}{2m}-\mu+V_{\rm ext}+g(n_0+2\tilde{n})\right] \Phi_0({\bf r}) \nonumber \\
        {}+g\tilde{m}\Phi_0^{*}({\bf r})=0,
\label{GP1}
\end{eqnarray}
where we introduced the condensate, the non-condensate, and the off-diagonal non-condensate densities as
\begin{eqnarray}
n_0({\bf r})&=&\mid\Phi_0\mid^2 , \\
\tilde{n}({\bf r})&=&\langle\tilde{\psi}^{\dagger}({\bf r})\tilde{\psi}({\bf r})\rangle ,\\
\tilde{m}({\bf r})&=&\langle\tilde{\psi}({\bf r})\tilde{\psi}({\bf r})\rangle ,
\end{eqnarray}
respectively. Now we employ the Popov ansatz \cite{Popov}:
\begin{eqnarray}
\tilde{m}({\bf r})&=&0,
\label{Popov}
\end{eqnarray}
which is known as a good approximation of a Bose gas at a finite temperature \cite{Giorgini}. In the Popov approximation, the equation for the condensate order parameter $\Phi_0({\bf r})$ (Eq.~(\ref{GP1})) is reduced to
\begin{eqnarray}
\left[-\frac{\hbar^2\nabla^2}{2m}-\mu+V_{\rm ext}+g(n_0+2\tilde{n})\right]\Phi_0({\bf r}) =0.
\end{eqnarray}
In a similar way, the equation for the non-condensate field operator $\tilde{\psi}({\bf r},t)$ is found to be

\begin{eqnarray}
i\hbar\frac{\partial \tilde{\psi}({\bf r},t)}{\partial t}&=&\biggl[-\frac{\hbar^2\nabla^2}{2m}+V_{\rm ext}-\mu\biggr]\tilde{\psi}({\bf r},t) \nonumber \\
           & & {}+2gn({\bf r})\tilde{\psi}({\bf r},t) \nonumber \\
           & &  {}+2g\Phi_0^2({\bf r})\tilde{\psi}^{\dagger}({\bf r},t).
\label{EX1}
\end{eqnarray}
The equation for $\tilde{\psi}^{\dagger}({\bf r},t)$ is the Hermitian conjugate of Eq.~(\ref{EX1}). This pair of equations is solved by expressing $\tilde{\psi}({\bf r},t)$ in the form

\begin{eqnarray}
\tilde{\psi}({\bf r},t)=\sum_{j}\left[u_j({\bf r})\hat{\alpha}_je^{-iE_j t/\hbar}-v_j^*({\bf r})\hat{\alpha}^{\dagger}_je^{iE_j^*t/\hbar}\right],
\label{BG1}
\end{eqnarray}
where $\hat{\alpha}_j$ and $\hat{\alpha}^{\dagger}_j$ are the annihilation and the creation operators satisfying the Bose commutation relation. Substituting Eq.~(\ref{BG1}) and its Hermitian conjugate into Eq.~(\ref{EX1}), we obtain the equation for the quasi-particle amplitudes $u_j$ and $v_j$: ({\it Bogoliubov-de Gennes equation})

\begin{eqnarray}
\left(
\begin{array}{cc}
\hat{L} & -g \Phi_0^2 \\
g \Phi_0^{*2} & -\hat{L}
\end{array}
\right)
\left(
\begin{array}{c}
u_j \\
v_j
\end{array}
\right)
=
E_j
\left(
\begin{array}{c}
u_j \\
v_j
\end{array}
\right),
\end{eqnarray}
where $\hat{L}$ is defined as
\begin{eqnarray}
\hat{L}=-\frac{\hbar^2 \nabla^2}{2m}+V_{\rm ext}-\mu+2gn({\bf r}).
\end{eqnarray}
The local density of the excitations $\tilde{n}$ can be expressed in terms of the quasi-particle amplitudes as
\begin{eqnarray}
\tilde{n}({\bf r})=\sum_{j} \left[ f_j (\mid u_j\mid^2+\mid v_j\mid^2)+\mid v_j\mid^2 \right],
\label{tilden}
\end{eqnarray}
where $f_j$ is the Bose distribution function of the $j$th excitation 
\begin{eqnarray}
f_j=\langle\hat{\alpha}_j^{\dagger}\hat{\alpha}_j\rangle=\frac{1}{e^{\beta E_j}-1}.
\end{eqnarray}

Now we are ready to introduce the Landau damping rate. Consider a low-lying collective mode (quasi-momentum $p$) of the condensate (quasi-momentum $k$) with energy $E_{k,p}$, surrounded by a static thermal cloud. The thermal cloud can absorb (or emit) quanta of collective modes of the condensate, leading a damping of the collective oscillation. Here we focus on the interaction between a collective mode of the condensate and a quasiparticle of the thermal cloud; the collective mode $p$ is coupled with the thermal excitation $i$, and transformed into the thermal excitation $j$, or {\it vice versa}. This process is known as the Landau damping process and it has been studied for current-free cases \cite{Giorgini, HM, SK, Liu, Pita1, Fed, Guilleumas, Jackson, Tsuchiya}. (There is another damping process known as Beliaev damping process; however, because its contribution is much smaller than the Landau damping at finite temperatures \cite{Giorgini}, we do not consider the Beliaev damping here.)

The Landau damping rate is calculated by perturbation theory \cite{Pita1,Giorgini} and given by

\begin{eqnarray}
\gamma_{k,p} = 4 \pi g^2 \sum_{i,j} |A_{i,j}|^2 (f_i-f_j) \delta(E_{k,p}+E_i-E_j).
\label{gammak}
\end{eqnarray}
Since we consider the current-carrying case, $A_{i,j}$ is a slight generalization of the expression presented in Refs.~\cite{Pita1,Giorgini}: 
\begin{eqnarray}
A_{i,j} &=& \int d{\bf r} \bigg[ \Phi_0^*\{u_p (u_iu^*_j+v_iv^*_j)+v_pu_iu^*_j\} \nonumber \\
        & & {} +\Phi_0\{u_p (u_iu^*_j+v_iv^*_j)+u_pv_iu^*_j\}\bigg].
\label{A}
\end{eqnarray}
Using these general expressions, we derive the damping rate for a Bose gas moving in an optical lattice in Sec.~\ref{ld1}.

\section{\label{ld1d}Landau damping in a dilute Bose gas moving in an optical lattice}
\subsection{\label{ft-tb}Finite-temperature tight-binding theory}

In this subsection, we develop a finite-temperature Bogoliubov theory for a Bose gas moving in a 1D optical lattice within a tight-binding approximation. We consider a Bose gas in a cigar-shaped magnetic trap combined with a periodic optical lattice. The radial trapping frequency is assumed to be large enough to ignore the motion in this direction. In addition, we assume that the longitudinal trapping frequency is negligibly small so that we ignore the effect of the magnetic trap in the longitudinal direction. As performed in recent experiments \cite{Fallani,Sarlo}, we consider the optical lattice moving in the longitudinal direction at a constant velocity $-v$. In order to properly apply the argument developed in Sec.~\ref{LD}, we employ the reference system which moves along with the optical lattice at the constant velocity $-v$. In this reference system, the superfluid part of the Bose gas {\it flows} with the velocity $v$, while the lattice potential is at rest and expressed as      

\begin{eqnarray}
V(z)=sE_R \sin\left(\frac{\pi z}{d}\right),
\end{eqnarray}
where 
\begin{eqnarray}
E_R=\frac{\hbar^2\pi^2}{2md^2}
\end{eqnarray}
is the photon recoil energy, $s$ is the dimensionless parameter, and $d$ is the lattice distance. We assume that $2k_BT/E_R\ll s$ so that the energy gap between the first and second excitation band is much larger than the temperature, and thus the first band is thermally occupied. We also assume that the optical lattice well is deep enough so that the wave functions are well localized on each lattice cite, allowing for using the tight-binding approximation \cite{Tsuchiya, Smerzi, Kramer}:

\begin{eqnarray}
\hat{\psi}(z,t)=\sum_{\ell}[\phi_{0\ell}+\tilde{\phi}_{\ell}(t)]f(z-\ell d),
\label{TB}
\end{eqnarray}
where 
$f(z-\ell d)$ is a localized function at the $\ell$th site, which satisfies the orthonormal relation:

\begin{eqnarray}
\int dz f(z-\ell d)f(z-md)=\delta_{\ell m},
\end{eqnarray}
where $\delta_{\ell,m}$ is the Kronecker's delta. With $f(z-\ell t)$, $\phi_{0\ell}$ and $\tilde{\phi}_{\ell}(t)$ form the condensate order parameter and non-condensate field operator at $\ell$th site respectively.

The following formalism is an analogue of the argument developed in Sec.~\ref{LD}. Substituting Eq.~(\ref{TB}) into Eq.~(\ref{EQM1}), and using the Popov ansatz~(\ref{Popov}), we obtain the discrete Gross-Pitaevskii equation including the mean-field of non-condensate parts

\begin{eqnarray}
-J(\phi_{0 \ell+1}+\phi_{0 \ell-1})+U(n_{0\ell}+2\tilde{n}_{\ell})\phi_{0 \ell} \nonumber \\
-\mu \phi_{0 \ell}=0,
\label{DNGP}
\end{eqnarray}
where condensate densities $n_{0\ell}$, non-condensate densities $\tilde{n}_{\ell}$, the hoping matrix elements $J$, and the on-site interaction energy $U$  are defined as 
\begin{eqnarray}
n_{0\ell}&=&\mid\phi_{0\ell}\mid^2 , \\
\tilde{n}_{\ell}(z)&=& \langle\tilde{\phi}_{\ell}^{\dagger}(z)\tilde{\phi}_{\ell}(z)\rangle ,
\end{eqnarray}
\begin{eqnarray}
J=-\int dz f_{\ell}\bigg[-\frac{\hbar^2}{2m} \frac{d^2}{d z^2}+V_{ext}(z)\bigg]f_{\ell+1},
\end{eqnarray}
\begin{eqnarray}
U=\tilde{g}\int dz \mid f_{\ell}\mid^4,
\end{eqnarray}
respectively. Here we introduced $\tilde{g}=2\hbar^2a/{m a_{\bot}^2}$ as the coupling constant modified for our effective 1D system using the radial length $a_{\bot}$ \cite{PandS}.
\begin{figure}[tb]
\epsfxsize=0.25\textwidth
\begin{center}
\includegraphics*[width=8.0cm,bb=0 0 300 200]{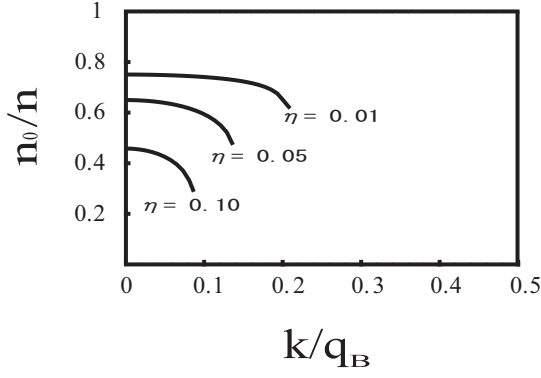}
\end{center}
\vspace*{-2mm}
\caption{Condensate fraction $n_0/n$ in an optical lattice as a function of the condensate quasimomentum $k/q_B$ at different temperatures $\eta=k_{B}T/J$, where $q_B=\pi/d$ is the Bragg vector and $U/J$ is fixed to $1$.}
\label{nc}
\end{figure}

Eq.~(\ref{DNGP}) admits the solution of the Bloch form (We set $\hbar=1$ from now on.)

\begin{eqnarray}
\phi_{0\ell}=\sqrt{n_{0\ell}}e^{ik\ell d}, 
\label{phik}
\end{eqnarray}
where $k$ is the quasimomentum of the condensate.
Substituting Eq.~(\ref{phik}) into Eq.~(\ref{DNGP}), one can write the chemical potential $\mu$ and the order parameter $\Phi_0$ as

\begin{eqnarray}
\mu=-2J\cos(kd)+U\left[n_0(T)+2\tilde{n}_0(T)\right] ,
\end{eqnarray}
and
\begin{eqnarray}
\Phi_0(z)&=&\sum_{\ell}\sqrt{n_{0\ell}}e^{ik\ell d}f(z-\ell d) \nonumber \\
         &\simeq&\sqrt{n_0}\sum_{\ell}e^{ik\ell d}f(z-\ell d).
\label{phi0}
\end{eqnarray}
where $n_0$ is the mean number of the condensate particle at a lattice cite.
Normal mode solutions are found by expanding $\tilde{\phi}_{\ell}(t)$ as 
\begin{eqnarray}
\tilde{\phi}_{\ell}(t)&=&\sum_{p}\bigg[\tilde{u}_{k,p}^\ell \hat{\alpha}_p e^{ipld-iE_p t} \nonumber \\
& &{}-\tilde{v}_{k,p}^{\ell *}\hat{\alpha}_p^\dagger e^{-ipld+iE_p^*t}\bigg]e^{ikld} .
\label{Bt}
\end{eqnarray}
Substituting Eq.~(\ref{Bt}) and its Hermitian conjugate into Eq.~(\ref{DNGP}), we obtain a pair of equations 

\begin{eqnarray}
\Big[2J\sin(kd)\sin(pd)+4J\cos(kd)\sin^2(pd/2) \nonumber \\
{}+Un_0 \Big] \tilde{u}_{k,p}^\ell-U\phi_{0\ell}^2\tilde{v}_{k,p}^\ell = E_{k,p}\tilde{u}_{k,p}^\ell, \\
\Big[2J\sin(kd)\sin(pd)- 4J\cos(kd)\sin^2(pd/2) \nonumber \\
{}-Un_0 \Big] \tilde{v}_{k,p}^\ell+U\phi_{0\ell}^{*2}\tilde{u}_{k,p}^\ell = E_{k,p}\tilde{v}_{k,p}^\ell,
\end{eqnarray}
where the excitation energy $E_{k,p}$ is 

\begin{eqnarray}
E_{k,p}=2J\sin(kd)\sin(pd)+\tilde{E}_{k,p}, 
\label{En}
\end{eqnarray}
with
\begin{eqnarray}
\tilde{E}_{k,p}&=&2\bigg[4J^2\cos^2(kd)\sin^4\bigg(\frac{pd}{2}\bigg) \nonumber \\
               & & {}+2JUn_0\cos(kd)\sin^2\bigg(\frac{pd}{2}\bigg)\bigg]^{\frac{1}{2}} .
               \label{tildeEn}
\end{eqnarray}
The Bogoliubov amplitudes $(u,v)$, which satisfy the normalization condition
\begin{eqnarray}
\int dz \big[\mid u_p\mid^2-\mid v_q\mid^2\big]=\delta_{p,q},
\end{eqnarray}
are expressed in Bloch forms as
\begin{eqnarray}
u_{k,p}(z)&=&\frac{1}{\sqrt{I}}\sum_\ell \tilde{u}^\ell_{k,p}f(z-\ell d) \nonumber \\
          &=&\frac{1}{\sqrt{I}}U_{k,p}\sum_{\ell}e^{i(p+k)\ell d}f(z-\ell d)
\label{u}
\end{eqnarray}
and
\begin{eqnarray}
v_{k,p}(z)&=&\frac{1}{\sqrt{I}}\sum_\ell\tilde{v}^\ell_{k,p}f(z-\ell d) \nonumber \\
          &=&\frac{1}{\sqrt{I}}V_{k,q}\sum_{\ell}e^{i(p-k)\ell d}f(z-\ell d).
\label{v}
\end{eqnarray}
In these expressions, $I$ is the number of lattice cites, while $U_{k,p}, V_{k,p}$ are given by
\begin{eqnarray}
U_{k,p}=\sqrt{\frac{1}{2}\left[\frac{4J\cos(kd)\sin^2(\frac{pd}{2})+Un_0}{\tilde{E}_{k,p}}+1\right]}Hê
\end{eqnarray}
\begin{eqnarray}
V_{k,p}=\sqrt{\frac{1}{2}\left[\frac{4J\cos(kd)\sin^2(\frac{pd}{2})+Un_0}{\tilde{E}_{k,p}}-1\right]}.
\end{eqnarray}
These satisfy the normalization condition
\begin{eqnarray}
\mid U_{k,p}\mid^2-\mid V_{k,p}\mid^2=1,
\label{normal}
\end{eqnarray}
{\it if $E_{k,p}$ is a real function}. We note that $\mid U_{k,p}\mid^2-\mid V_{k,p}\mid^2=0$ if $E_{k,p}$ contains an imaginary component. This problem was first pointed out by Niu and Wu \cite{Niu}. We confirm that it is still valid in our tight-binding model. We will discuss this important issue again in Sec.~\ref{inst} regarding the instability of condensates.

The number of the condensate at a lattice site, $n_0$, is determined by the relation $n = n_0+\tilde{n}$, where $\tilde{n}$ is given in Eq.~(\ref{tilden}). Note that $n_0$ implicitly depends both on the temperature $T$ and the condensate quasimomentum $k$.  Fixing parameters $U/J=1$, $n=2$ and $I=250$, we solved the equation $n = n_0+\tilde{n}$ self-consistently. In Fig.~\ref{nc}, the condensate fraction $n_0/n$ at different temperatures $\eta=k_BT/J$ is shown as a function of $k$. Due to the fatal limitation of the Bogoliubov theory, one cannot determine $n_0$ at $k\ge k_c$, where $k_c$ is the critical quasimomentum determined by Landau criterion of superfluidity \cite{Niu}, since the low-energy excitation becomes negative. We will give the explicit expression of $k_c$ in Sec.~\ref{CC}, after showing in Sec.~\ref{LLsub} that $k_c$ turns out to be the same as the critical quasimomentum determined from the Landau damping rate. 


\subsection{\label{ld1}Landau damping in a moving 1D optical lattice}

\begin{figure}[bt]
\begin{center}
\includegraphics*[width=8.0cm,bb=0 0 470 430]{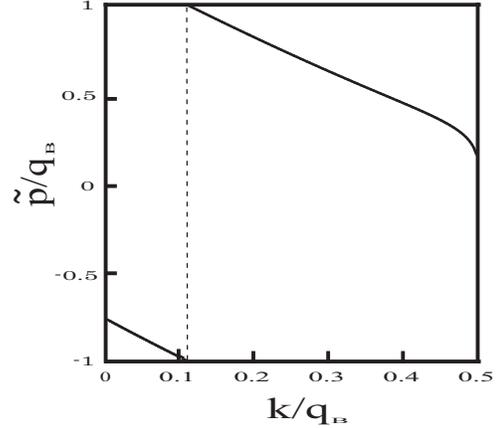}
\end{center}
\vspace*{-2mm}
\caption{Quasimomentum $\tilde{p}$ satisfying $E_p+E_{p+\tilde{p}}=E_{\tilde{p}}$ with $p \ll \tilde{p}$. Both $\tilde{p}$ and $k$ are normalized by the Bragg vector $q_B=\pi/d$, where $d$ is a lattice spacing. $\alpha=Un_0/J$ is set to $1$.}
\label{tildep}
\end{figure}

Now we calculate the Landau damping rate for a Bose gas in a moving 1D optical lattice using the tight-binding approximation introduced in Sec.~\ref{ft-tb}. First, we calculate the matrix element $A_{i,j}$ in Eq.~(\ref{A}). By using Eqs.~(\ref{phi0}), (\ref{u}) and (\ref{v}), we obtain,

\begin{eqnarray}
A_{i,j}=\sqrt{\frac{n_0}{I}}\frac{U}{g}\delta_{p+q_i,q_j+G_n}\Big[ U_p(U_iU_j+V_iV_j)\nonumber \\
{}+V_pU_iU_j +V_p(U_iU_j+V_iV_j)+U_pV_iU_j \Big],
\label{A'}
\end{eqnarray}
where $G_n=2n\pi/d$ (n is an integer) is the reciprocal lattice vector corresponding to the Umklapp scattering process when $n$ is not zero. We see that Eqs.~(\ref{gammak}) and (\ref{A'}) involves the momentum and energy conserving delta functions. This means that, in order for the Landau damping to occur, there must exist $q_i$ and $q_j$ that satisfy the energy and momentum conservation \cite{Tsuchiya}. Combining the two conservations, we require that there exists $q$ that satisfies 

\begin{eqnarray}
E_{k,p}+E_{k,q}=E_{k,p+q+G_n}.
\label{require}
\end{eqnarray}
For small quasimomentum $pd \ll  1$, $E_{k,p}$ in Eq.~(\ref{En}) has a linear dispersion in terms of $p$; namely  
\begin{eqnarray}
E_{k,p}=[-2J\sin(kd)+\sqrt{2JUn_0\cos(kd)}]\mid p\mid,
\label{Ep}
\end{eqnarray}
where we set the $k$'s direction as positive and considered the energy dispersion with $p<0$, since only this type of excitation is responsible for the energetic instability, as we show in Sec.~\ref{LL}.

Assuming that $p$ is small enough that $E_{k,p}$ can be expressed as Eq.~(\ref{Ep}), one can reduce required the requiring condition Eq.~(\ref{require}) reduces to 
\begin{eqnarray}
[2J\sin(kd)-\sqrt{2JUn_0\cos(kd)}]=\frac{\partial E_{\tilde{p}}}{\partial (\tilde{p}d)},
\label{condition}
\end{eqnarray}
where we denoted $\tilde{p}=q+G_n$. Note that the Umklapp scattering is a mere superficial problem for us here, since $\tilde{p}$ only appears in the periodic energy expressions in Eqs.~(\ref{En}) and (\ref{tildeEn})).

Fig.~\ref{tildep} shows $\tilde{p}$ in Eq.~(\ref{condition}) as a function of $k$ with $\alpha\equiv Un_0/J=1$. It is seen that the Umklapp scattering appears at a finite-$k$, while it cannot be seen when $k=0$ as shown in Ref.~\cite{Tsuchiya}. 

\begin{figure}[bt]
\begin{center}
\includegraphics*[width=8.0cm,bb=0 0 432 340]{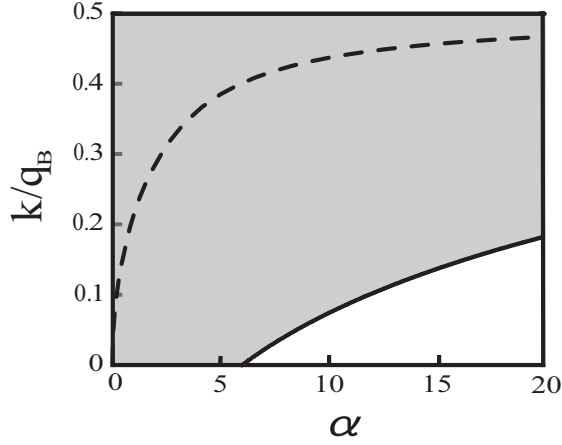}
\end{center}
\vspace*{-2mm}
\caption{Damping possibility diagram. Damping processes are possible to exist only in the shaded area. Dashed line indicates the critical quasimomentum of the condensate given by Eq.~(\ref{kc}).}
\label{possible}
\end{figure}

The condition for $\tilde{p}$ to exist, determined by Eq.~(\ref{condition}), is shown in Fig.~\ref{possible}. At $k=0$, all the Landau damping processes disappears when $\alpha>6$, which agrees with the result derived in Ref.~\cite{Tsuchiya}. With increasing $k$, we find that the region where the Landau damping process disappears prominently decreases. This disappearance can be confirmed in future experiments.

By integrating Eq.~(\ref{gammak}) using the delta-function for the energy conservation, we explicitly obtain the expression of the Landau damping rate for a Bose gas moving in a 1D optical lattice:
\begin{eqnarray}
\gamma_{k,p} &=& \frac{E_{k,p}}{\mid\partial^2E_{k,\tilde{p}}/\partial^2\tilde{p}\mid}\frac{U\beta d^2\sqrt{JUn_0\cos(kd)}}{4\sinh^2\left(\frac{\beta E_{k,\tilde{p}}}{2}\right)} \nonumber \\ 
             & & \times\Biggl[\frac{8J\cos(kd)\sin^2(\frac{\tilde{p}d}{2})+Un_0}{2\tilde{E}_{k,\tilde{p}}} \nonumber \\
             & &   {}- \frac{J(Un_0)^2\cos(kd)\sin(\tilde{p}d)}{\tilde{E}_{k,\tilde{p}}^2 \sqrt{JUn_0\cos(kd)}}\Biggr]^2.
\label{gammap}
\end{eqnarray}

\section{\label{LL}Landau damping and Landau instability}
\subsection{\label{LLsub}Landau damping and Landau instability}

\begin{figure}[tb]
\epsfxsize=0.25\textwidth
\begin{center}
\includegraphics*[width=8.0cm,bb=0 0 500 750]{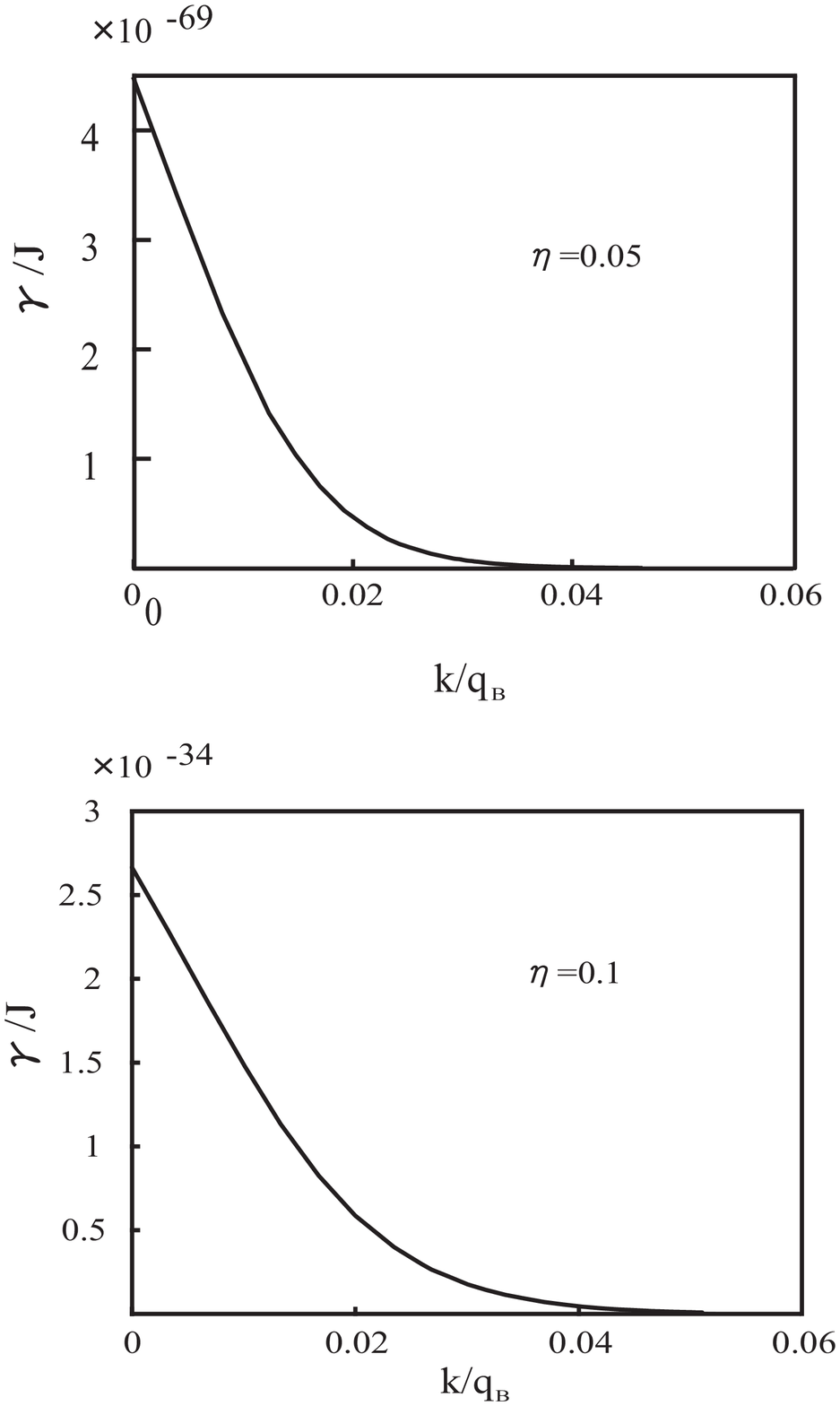}
\end{center}
\vspace*{-2mm}
\caption{Landau damping rate $\gamma_{k,-0.01q_B}$ in the 1D optical lattice with $U/J=1$ as a function of a condensate quasimomentum $k < k_c$ at different temperatures $\eta=k_B T/J$. 
$\gamma$ and $k$ are normalized to the hopping matrix element $J$ and the Bragg vector $q_B$ respectively.} 
\label{gamma}
\end{figure}

In this section, we show that the Landau damping process is closely related to the breakdown of superfluidity due to the energetic instability (or Landau instability).
As explicitly shown in Ref.~\cite{Giorgini}, the amplitude of the collective mode $\delta\Phi_{k,p}$ decays (or grows) as 
\begin{eqnarray}
\delta\Phi_{k,p} \propto e^{-\gamma_{k,p}t}, 
\label{delta}
\end{eqnarray}
where $\gamma_{k,p}$ is the Landau damping rate discussed in this paper.
This relation indicates that the condensate with a superfluid current $k$ is stable as long as $\gamma_{k,p}$ is positive for any collective mode momentum $p$. In fact, in a uniformly resting ($k=0$) condensate, the Landau damping rate is always positive.  Thus an induced collective mode decays exponentially in time \cite{Pita1, Giorgini}. This means that the thermally populated cloud stabilizes the condensate part. However, as shown below, it is possible that $\gamma_{k,p}$ becomes negative for a moving condensate with finite-$k$. Negative $\gamma_{k,p}$ indicates the appearance of the inverse process of the usual Landau damping process, whereby the thermal clouds emit excitation quanta. The increase of the amplitude of collective modes severely destabilizes the condensate state, leading to the breakdown of the superfluidity. 

We now show that the Landau damping rate in Eq.~(\ref{gammap}) changes its sign at a critical quasimomentum.
We find that the sign of $\gamma_{k,p}$ fully depends on the sign of $E_{k,p}$, {\it i.e.}
\begin{eqnarray}
{\rm sgn}(\gamma_{k,p})={\rm sgn}(E_{k,p}).
\label{sgn}
\end{eqnarray} 
This means that the critical quasimomentum of the Landau damping rate coincides with the critical quasimomentum of the corresponding collective mode determined by the Landau criterion, which is the Bogoliubov sound velocity when we consider a dilute Bose gas in a uniform system. Landau originally argued that the superfluid decays by emitting phonon, whose energy level is lower than the superfluid state, which is indeed possible when the superfluid possess a velocity larger than the critical velocity ({\it Landau criterion}) \cite{Landau}. Here we find that the critical velocity obtained from the Landau damping rate coincides with the velocity derived from the Landau criterion, meaning that the Landau damping process is the microscopic mechanism of the macroscopic Landau instability. 

This close relation between the Landau damping and the Landau instability (Eq.~(\ref{sgn})) can also be derived from the general expression of the Landau damping rate in Eq.~(\ref{gammak}). The sign of $\gamma_{k,p}$ in Eq.~(\ref{gammak}) is determined by the difference $f_i - f_j$ between the Bose distribution functions of static thermal clouds. Combining this with the energy conserving delta-function, we find that $\gamma_{k,p}$ becomes negative when $E_{k,p}$ becomes negative (Eq.~(\ref{sgn})). This means that the Landau damping process is generally the source of the Landau instability within the approximation of static thermal clouds. 

Fig.~\ref{gamma} shows the Landau damping rate at $k<k_c$ at different temperatures $\eta=k_B T/J$. The damping rate prominently decreases with increasing $k$, meaning that the stabilization mechanism of the condensate by the surrounding thermal cloud (usual Landau damping) works less effectively at a larger $k<k_c$.  

Although we do not show the growth rate $\gamma_{k,p}$ of an unstable mode at $k>k_c$ due to the fact that the condensate density at the region cannot be calculated in our Bogoliubov theory, we estimate its behaviour as follows. As suggested by the experiment in Ref.~\cite{Sarlo}, assuming that the condensate density remains constant for $k>k_c$ to its minimum value shown in Fig.~\ref{nc}, we confirmed that $\gamma_{k,p}$ prominently decreases with increasing $k>k_c$ and diverges to $-\infty$ at the edge $k=q_B /2$. Moreover, the decreasing rate of $\gamma_{k,p}$ is confirmed to be the increasing function of the temperature within the assumption. This indeed means that the amount of thermal clouds is crucially important in the breakdown process of superfluids. 

To summarize, we propose the following general scenario for the breakdown of superfluidity due to the Landau damping process with thermally populated excitations: As long as $k<k_c$, the Landau damping rate $\gamma_{k,p}$ is positive, and thus the superfluid state is stable. However, once $k$ exceeds $k_c$, the amplitude of collective modes increase exponentially in time by the inverse of process of the usual Landau damping process. This critical process should lead the breakdown of the superfluidity. On the other hand, if there is no thermal cloud, neither the Landau damping nor the collisional damping \cite{Konabe} can emerge. Thus the superfluid does not breakdown. 

In short, we theoretically proved that thermal excitations play a significant role in the breakdown of the superfluid due to the energetic instability.

\subsection{\label{CC}Critical current}

\begin{figure}[bt]
\begin{center}
\includegraphics*[width=8.0cm,bb=0 0 470 450]{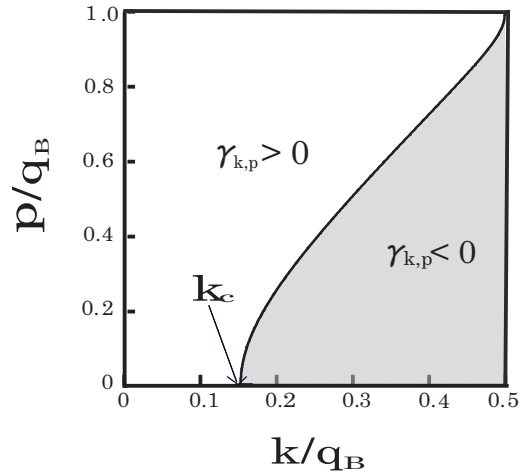}
\end{center}
\vspace*{-2mm}
\caption{Stability diagram of the BEC in the optical lattice with $\alpha=0.5$. The shaded area is where the excitation energy $E_{k,p}$ is negative thus the system is energetically unstable. In this area, the Landau damping rate $\gamma_{k,p}$ is also negative (Eq.~(\ref{sgn})) and the inverse process of the usual Landau damping is dominant. Both $k$ and $p$ are normalized by $q_B=\pi/d$, where $d$ is a lattice spacing.}
\label{unstable}
\end{figure}

We now evaluate the specific critical quasimomentum $k_c$ of the condensate, where the superfluid starts to decay due to the inverse process of the usual Landau damping for a dilute Bose gas moving in an optical lattice. The desired $k_c$ is obtained by requiring $\partial_p E_{k_c,p}=0$ in Eq.~(\ref{En}). Then, $k_c$ is expressed as a function of the quasimomentum $p$ of an excitation; however, the critical current of a superfluid should be the minimum current with which the superfluid starts to decay. As seen in the Fig.~\ref{unstable}, the smaller the quasimomentum of an excitation, the smaller the quasimomentum of the condensate required for negative excitation energy to appear. Thus the critical quasimomentum of the condensate should be the quasimomentum with which the energy of the phonon corresponding to the longest wave length becomes zero (see Fig.~\ref{unstable}). At this critical current, which is nothing but the Landau critical current, the superfluid state becomes unstable if the system contains enough amounts of thermal excitations.

Therefore, by equating Eq.~(\ref{condition}) to zero, we obtain

\begin{eqnarray}
\cos(k_c d)=\frac{-\alpha+\sqrt{\alpha^2+16}}{4},
\label{kc}
\end{eqnarray}
where $\alpha=Un_0/J$. We note that the Landau instability is found to take place only where the Landau damping process exists as seen in Fig.~\ref{possible}. 
This shows that the very close relation between Landau instability of condensates and the Landau damping process.   

We note that our 1D results can be tested in the definitive 1D optical lattice system using the recent technique \cite{Morsch, Moritz, Stoferle}, which realizes a much tighter (in the radial direction) magnetic trap than used in Ref.~\cite{Sarlo}.   
Experiments should also be conducted in the collisionless regime.  


\section{\label{inst}Landau instability and Dynamical instability in an optical lattice}

In this paper, we showed that the Landau damping works as a source of Landau instability at a finite temperature. The thermal cloud around the superfluid tempts the condensate into exciting by accelerating an energetically irresistible process, {\it i.e.} the inverse process of the usual Landau damping. This explains the microscopic mechanism of the Landau instability at a finite temperature. 

There is another well known instability of the condensate in an optical lattice, the dynamical instability \cite{Niu, Menotti, Demler}. In this section, we briefly discuss how the two kinds of instabilities are understood in connection with our results. 

\subsection{Similarities}

As described in Sec.~\ref{LL}, a collective mode $\delta\Phi_{k,p}$ of the condensate grows or decays in time as
\begin{eqnarray}
\delta\Phi_{k,p} \propto e^{-iE_{k,p}t-\gamma_{k,p}t}.
\end{eqnarray}
If $\gamma_{k,p}$ becomes negative, the corresponding collective mode oscillation exponentially grows, leading to the breakdown of the superfluid. This is Landau instability discussed in this paper. On the other hand, it is also possible in some systems that $E_{k,p}$ itself contains an imaginary part. Then likewise, the excitation exponentially grows and the superfluid severely decays. This is the dynamical instability, which was characterized by the exponential growth of collective modes in time in former literatures \cite{Niu, Menotti}. However, as seen in our argument, Landau instability can also be related to the exponential growth of collective modes. 

\subsection{Differences: collective modes and elementary excitations}

One obvious difference between Landau instability and the dynamical instability is the scales of the critical velocity, at least for a Bose gas in an optical lattice discussed in this paper. As seen in Eq.~(\ref{En}), Landau instability takes place at $0<k<q_B/2$ while the dynamical instability occurs at $q_B/2<k$, where $E_{k,p}$ contains an imaginary component.

However, we note here that the most significant difference between Landau instability and the dynamical instability is weather it is caused by modulational {\it collective modes} or quantum {\it elementary excitations}. We use the {\it collective mode} for the ``classical" modulation of the condensate itself, while the {\it elementary excitation} is used to mean the ``quantum" and thermal depletion of the non-condensate part \cite{PandS}.  

As seen in Eq.~(\ref{normal}), the Bogoliubov amplitudes $(U,V)$ satisfy the usual normalization condition as long as the corresponding energy $E_{k,p}$ in Eq.~(\ref{En}) is a real function. Then $(U,V)$ can be sufficiently quantized and $E_{k,p}$ can be regarded as the energy of an {\it elementary excitation} \cite{PandS}.
However, once $E_{k,p}$ contains an imaginary part, or equivalently $\tilde{E}_{k,p}$ becomes purely imaginary, $(U,V)$ cannot satisfy the normalization condition, but instead (with some straightforward algebra) 
\begin{eqnarray}
\mid U_{k,p}\mid^2-\mid V_{k,p}\mid^2=0,    
\end{eqnarray}
which is also obtained using an abstract model in Ref.~\cite{Niu}.
This means that $(U,V)$ {\it cannot} be quantized, thus they should be regarded as the amplitudes of the {\it collective modulation} of the condensate.

In summary, Landau instability is caused by the thermal {\it elementary excitation} while the dynamical instability is caused by the {\it collective modulation} of the condensate itself. Consequently, the Landau instability only occurs at finite temperatures while the dynamical instability can be induced even at $T=0$. This agrees with our analysis on the Landau damping that there should be a thermal cloud in order to induce the Landau instability since the thermal cloud is also a vital source for the Landau damping process. As noted in Sec.~\ref{intro}, this temperature dependence of the decay of superfluids is experimentally observed in Ref.~\cite{Sarlo}, which is consistent with our analysis on the {\it elementary excitations} and {\it collective modes}. We hope that our work will stimulate further experiments on the breakdown of superfluids.

\section{\label{con}conclusion}

In summary, we investigated the instability of a moving condensate in an optical lattice surrounded by thermally excited quasiparticles at finite temperatures. We revealed that the thermal clouds destabilize the condensate by the inverse process of the usual Landau damping. 
Also, we explicitly calculated the Landau damping rate for a Bose gas at a finite temperature and showed the relation between Landau damping and Landau instability. We found a certain condition for energy scales that Landau damping or Landau instability occurs. We finally noted the difference and similarity between the Landau instability and the dynamical instability in an optical lattice.

\begin{acknowledgments}
We wish to thank K. Kamide, N. Yokoshi, D. Yamamoto, S. Kurihara, and J. E. Williams for valuable discussions. Also, H. Shibata's technical assistance in numerical calculations is greatly acknowledged. S.K. and I.D. are supported by JSPS Research Fellowship for Young Scientists. 
\end{acknowledgments}


\end{document}